\documentclass[conference]{IEEEtran}
\usepackage{graphicx} 
\usepackage{tabularx}
\usepackage{enumitem}
\usepackage{listings}
\usepackage{xcolor}
\usepackage{array}

\newlist{prlist}{enumerate}{1}
\setlist[prlist]{%
  label={[P\arabic*]},      
  ref   =P\arabic*,         
  leftmargin=*,
  font  =\scriptsize             
}

\usepackage[many]{tcolorbox}

\newtcolorbox{fancybox}[1][]{
  enhanced,
  attach boxed title to top center={yshift=-3mm,yshifttext=-1mm},
  colback=blue!5!white,
  colframe=blue!25!black,
  colbacktitle=blue!25!black,
  fonttitle=\bfseries,
  title=#1,
  boxed title style={size=small, colframe=blue!45!black, colback=blue!45!white},
  drop fuzzy shadow,
  width=\textwidth,
  breakable
}

\newcommand\CRV[1]{{\textcolor{black}{#1}}}

\lstdefinestyle{prompt}{
    basicstyle=\scriptsize,
    backgroundcolor=\color{gray!10},
    frame=single,
    breaklines=true,
    keepspaces=true,
    columns=fullflexible,
    captionpos=b
}

\newcommand{\sectopic}[1]{\vspace{0.2em}\par\noindent{\textit{\bfseries #1}}}

\title{Prompt Engineering for Requirements Engineering: A Literature Review and Roadmap}
\author{}
\author{
    \IEEEauthorblockN{Kaicheng Huang\IEEEauthorrefmark{1}, Fanyu Wang\IEEEauthorrefmark{1}, Yutan Huang\IEEEauthorrefmark{1}
    Chetan Arora\IEEEauthorrefmark{1}
  }
    \IEEEauthorblockA{\IEEEauthorrefmark{1}Faculty of Information Technology, Monash University, Melbourne, Australia}
    Email: 
    chetan.arora@monash.edu
\vspace*{-1em}
}
\date{April 2025}

\begin{document}
\maketitle

\begin{abstract}
\CRV{Advancements in large language models (LLMs) have led to a surge of prompt engineering (PE) techniques that can enhance various requirements engineering (RE) tasks. However, current LLMs are often characterized by significant uncertainty and a lack of controllability. This absence of clear guidance on how to effectively prompt LLMs acts as a barrier to their trustworthy implementation in the RE field.} We present the first roadmap-oriented systematic literature review of Prompt Engineering for RE (PE4RE). Following Kitchenham’s and Petersen’s secondary-study protocol, we searched six digital libraries, screened 867 records, and analyzed 35 primary studies. To bring order to a fragmented landscape, we propose a hybrid taxonomy that links technique-oriented patterns (e.g., few-shot, Chain-of-Thought) to task-oriented RE roles (elicitation, validation, traceability). Two research questions—with five sub-questions—map the tasks addressed, LLM families used, and prompt types adopted, and expose current limitations and research gaps. Finally, we outline a step-by-step roadmap showing how today’s ad-hoc PE prototypes can evolve into reproducible, practitioner-friendly workflows.



\end{abstract}

\section{Introduction}

Requirements Engineering (RE) is fundamental to the success of every software project~\cite{hofmann2001requirements}. With the rapid advancements in large language models (LLMs) and other natural language processing (NLP) technologies, NL specifications have become increasingly common. These specifications can be found in user stories within agile backlogs, regulatory clauses in safety-critical domains, and free-form stakeholder comments stored in issue trackers. Their growing prevalence encourages using NLP techniques in various RE tasks, including elicitation, classification, conflict detection, traceability-link recovery, and even automated test generation~\cite{arora2023advancing}.

Due to the complexity of state-of-the-art LLMs, which contain hundreds of billions of parameters, most researchers in RE primarily access these models through hosted inference APIs rather than fine-tuning the entire models. \CRV{Since LLMs are known for uncertainty and uncontrollability~\cite{hou2024enhancing,zhang2025automatically}, regular users may find inconsistency when performing inference jobs via API services, which challenges the trustworthiness of LLMs usage~\cite{sahoo2024systematic,fagbohun2024empirical}.} This shift in access changes the focus from model architecture to prompt engineering (PE)—the art of crafting input queries that guide an LLM toward producing accurate, consistent, and context-appropriate outputs~\cite{marvin2023prompt}. 

PE techniques include methods such as few-shot priming~\cite{brown2020language}, chain-of-thought reasoning~\cite{wei2022chain}, retrieval-augmented generation~\cite{lewis2020retrieval}, constraint injection, and multi-role dialogue orchestration. These techniques have become essential for applying LLM capabilities within the RE community. 
However, there is still a lack of comprehensive understanding of the full range of PE techniques and their trade-offs in the RE community. As a result, researchers often treat prompting as an ad hoc implementation detail rather than a deliberate design choice with significant effects on outcomes.

Consequently, researchers might (i)~choose popular prompting patterns, such as ``Act as a requirements analyst,'' without verifying whether the framing aligns with relevant downstream artifacts (e.g., traceability matrices, compliance reports); (ii)~develop elaborate task-specific templates while neglecting to conduct ablation studies, which could reveal which elements of the prompt—such as examples, reasoning cues, or structural constraints—are responsible for any observed improvements; and (iii)~report isolated success metrics that obscure the practical limitations of their approach.
Without a unified framework mapping how PE strategies connect with RE tasks, the field struggles to determine which techniques are the most effective, where they fall short, or what realistic benefits practitioners can anticipate. 
To overcome the limitations, we proposed a guideline paper that systematically analyzes existing studies and provides a roadmap for future developers in this field. Our contributions are listed as follows:

\begin{itemize}
    \item Following Kitchenham et al.~\cite{kitchenham2022segress}’s guidelines for secondary studies and Petersen et al.~\cite{petersen2015guidelines}’s mapping methodology, we searched six major digital libraries, filtered 867 records down to 35 primary studies, and extracted detailed data about their LLM choice, prompt design, RE task, and more.
    \item We map existing studies into our proposed PE classification from both the feature and task perspectives and systematically discuss the advantages and disadvantages in the context of RE tasks.
    \item A comprehensive guideline is proposed for future researchers with the potential limitations and opportunities.
\end{itemize}

\sectopic{Structure.} Section~\ref{sec:background} describes the background of Requirements Engineering (RE) and Prompt Engineering (PE). Section~\ref{sec:methodology} specifies the methodology of our literature review. Section~\ref{sec:pe_cate} gives the categorization of PE techniques in the context of RE domain. Section~\ref{sec:results} presents our analysis results and the roadmap discussion. Section~\ref{sec:conclusion} concludes our study.
\section{Background}
\label{sec:background}
\subsection{Requirements Engineering}
Requirements Engineering (RE) systematically elicits, analyses, specifies, validates, and manages the requirements defining a software system’s scope and success~\cite{Sommerville2012InformationRF}. Acting as a conduit between stakeholder needs and technical implementation, RE negotiates diverse perspectives to produce verifiable requirements~\cite{Knutas2017AMD, Udousoro2020EffectiveRE}. Its iterative cycle, comprising elicitation, analysis, specification, validation, and ongoing management, ensures that new insights and contexts continually refine the requirements~\cite{Besrour2015ExploratoryST,wang2025requirementsdrivenautomatedsoftwaretesting}. High-quality RE reduces costly rework and drift, and techniques like LLM–assisted analysis and automated traceability are now reshaping the practice, accelerating processing while reinforcing RE's central goal: delivering the right software to the right users.

\subsection{Prompt Engineering}
Prompt engineering (PE) treats the prompt—the textual query to a large language model—as a tunable interface for controlling outputs, combining task instructions, domain context or examples, and the user’s input to maximise performance without altering model weights~\cite{Liu2021PretrainPA}. Since LLMs are accessed via hosted APIs rather than routinely fine-tuned, PE provides a cost-effective alternative for rapid domain adaptation~\cite{Jiang2025GenerativeRD}. Designers iteratively refine prompts—testing phrasing, example selection, and structure—to achieve precise, verifiable outputs~\cite{Minaee2024LargeLM}. With its own design patterns (e.g., few-shot priming, chain-of-thought, retrieval-augmented prompts) and taxonomies (open vs. closed, instructive vs.\ conversational, simple vs.\ composite), PE is now likened to a new programming paradigm~\cite{VelsquezHenao2023PromptEA}. Mastering PE is increasingly essential for deploying LLMs for safety and quality.

\section{Systematic Literature Review of PE in RE}
\label{sec:methodology}
\begin{figure}
    \centering
    \includegraphics[width=0.9\columnwidth]{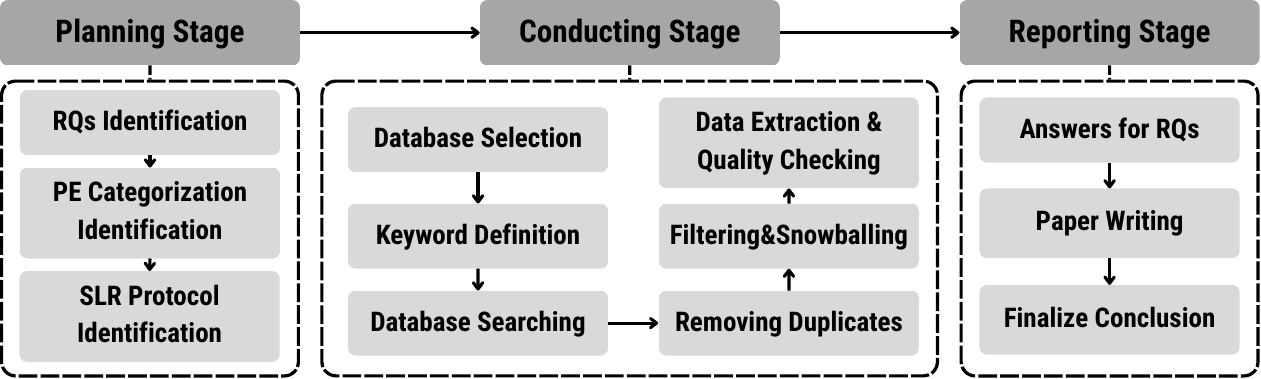}
    \vspace{-0.5em}
    \caption{Procedure of the Study}
    \label{fig:procedure}
    \vspace{-1.5em}
\end{figure}
In this section, we outline our systematic review methodology, including the search strategy and data extraction procedures. Our approach follows the guidelines proposed by Kitchenham et al.~\cite{kitchenham2022segress} for conducting systematic literature reviews (SLRs), and Petersen et al.~\cite{petersen2015guidelines} for conducting preliminary literature reviews, particularly in the context of our roadmap paper. There are three stages in our SLR, including (1) the \emph{planning} stage involves identifying research questions, particularly from a roadmap perspective, and outlining a detailed research plan for the study; (2) the \emph{conducting} stage involved analyzing and synthesizing existing primary studies to answer potential questions; (3) the \emph{reporting} stage involved analyzing, concluding, and discussing the PE4RE studies in context of advanced PE researches and identifying the challenges, limitations and future directions in this field. The overview of this work is presented in Fig.~\ref{fig:procedure}.

\subsection{Research Questions in Roadmap}
\sectopic{RQ1. How is the prompt technique currently engineered to support specific RE tasks?} Numerous studies have utilized PE methods to enhance the RE process. In this RQ, we will primarily focus on the technical aspects of prompting methods, specifically discussing (1) the large language models (LLMs) used and (2) the categorization of the prompting methods.
\begin{table}[]
\scriptsize
\centering
\caption{Search Strings}
\vspace{-0.5em}
\label{tab:search_strings}
\begin{tabularx}{\linewidth}{lX}
\hline
\textbf{Groups}     & \textbf{Search Strings}                                                                                                                                                                                                                 \\ \hline
\textbf{PE Group}   & ``prompt engineering'' OR ``prompt design*'' OR ``prompt creat*'' OR ``prompt construct*'' OR ``prompt pattern*'' OR ``prompt-based'' OR ``PE''                                                                                \\
\textbf{RE Group}   & ``requirement* engineering'' OR ``requirements analy*'' OR ``requirements specification*'' OR ``software requirement*'' OR ``requirements elicit*'' OR ``requirements valid*'' OR ``requirements manage*'' OR ``SE'' OR ``RE'' \\
\textbf{LLMs Group} & ``LLM'' OR ``Large Language Model*'' OR ``Language Model*'' OR ``LM*'' OR ``PLM*'' OR ``Pre-trained'' OR ``Pre-training'' OR ``Transformer'' OR ``generative AI'' OR ``GenAI'' OR ``generative artificial intelligence'' \\ \hline     
\end{tabularx}%
\vspace{-1em}
\end{table}

\sectopic{RQ2. What is the current limitation and future direction of PE4RE in view of the advanced PE technologies?} From PE (NLP domain) to RE (SE domain), we acknowledge the research and technical gap between these two different fields. In RQ2, we compare the current PE4RE research with the advanced PE studies to identify limitations or challenges, then we discuss the potential solutions and outline a roadmap of PE4RE research.

\subsection{Searching Procedure}
\subsubsection{Database Selection}
We selected six online databases for our survey, including IEEE Xplore, ACM Digital Library, Wiley, Scopus, Web of Science, and Science Direct. \CRV{These databases crossly achieve comprehensive coverage for RE studies, including RE journal, REFSQ conference, etc.} We acknowledge that there are other studies that have adopted Arxiv as their database. But considering that some of the studies on Arxiv have not been completely reviewed, we didn't include Arxiv in our database selection.


\begin{table}[]
\scriptsize
\caption{Inclusion and Exclusion Criteria}
\vspace{-0.5em}
\label{table:criteria}
\begin{tabularx}{\linewidth}{lX}
\hline
\textbf{ID}  & \textbf{Description}
\\ \hline
\multicolumn{2}{l}{\textbf{Selection Criteria}} \\ \hline
S01          & Papers written in English. \\
S02        & Papers in the subject of ``Computer Science". \\
S03          & Papers published from 2018 to the present. \\ 
S04          & Papers with accessible full text.
\\ \hline
\multicolumn{2}{l}{\textbf{Inclusion Criteria}} \\ \hline
I01 & Focuses on at least one RE task, i.e., requirements elicitation, analysis, specification, validation, or management in SE.
\\
I02 & Use prompt engineering strategies in the methodology – even if the paper does not explicitly use the term ``prompt engineering,'' these techniques should be evident in how the solution is implemented. \\ \hline
\multicolumn{2}{l}{\textbf{Exclusion Criteria}} \\ \hline
E01 & Studies that only mentioned prompt engineering but no details or guidelines. \\
E02 & Papers with fewer than 5 pages (1-4 pages). \\
E03 & Vision papers and grey literature (unpublished work), books (chapters), posters, discussions, keynotes, magazine articles, and comparison papers. \\ \hline
\end{tabularx}
\vspace{-1em}
\end{table}

\subsubsection{Searching Strings}
\CRV{To ensure comprehensive coverage, we defined three groups of search strings. We applied them, using wildcards (`*`) for term variations, `OR` within groups, and `AND` between groups, to the “Title, Keywords, and Abstract” fields (see Table~\ref{tab:search_strings}). The first and second authors iteratively refined each string, and then all four authors consolidated the final set. Although treating ``Prompt* Engineering'' as a mandatory group may have excluded studies that introduce novel prompting techniques without explicit mention; this approach maximises reproducibility and precision by selecting only those papers that discuss prompt engineering. }

\subsubsection{Selection Criteria}
We present our selection criteria in Table.~\ref{table:criteria}. Taking into account some similar expressions in the other field, we select the subject of ``computer science'' to avoid returning some irrelevant papers. As for \texttt{S03}, due to the first release of ``LLM'' or ``pretrained-LM'' in 2018, we applied this criterion to ensure the coverage. \CRV{This step is conducted by the first author, which yields 867 studies (before removing duplicates) and 538 studies (after removing duplicates).}

\subsubsection{Inclusion and Exclusion Criteria}
\CRV{After our database search, we applied the inclusion and exclusion criteria listed in Table~\ref{table:criteria}. Criterion \texttt{I01} excludes papers that do not address requirements engineering, and \texttt{I02} requires each paper to propose at least one prompting method in its methodology. First and second authors independently screened the titles and abstracts of 538 retrieved records, identifying 60 and 50 candidates, respectively, with overlap. Third and fourth authors then cross-validated these candidates and conducted a manual quality assessment, narrowing the list to 31 studies. Finally, snowballing added four more papers, yielding a total of 35.}

\begin{table}[]
\scriptsize
\centering
\caption{Details of Searching Results}
\vspace{-0.5em}
\label{tab:searching_results}
\resizebox{\columnwidth}{!}{
\begin{tabular}{ccccc}
\hline
Database      & Searching & Remove Duplicate & Independent Screening & Discussion and Quality Assessment                \\ \hline
ACM           & 191       & 174    &15(1st Aut.)/10(2st Aut.)          & 4                    \\
IEEE          & 134       & 134    &17(1st Aut.)/18(2st Aut.)         & 14                 \\
Wiley         & 2         & 2      &0(1st Aut.)/0(2st Aut.)          & 0                    \\
WOS           & 246       & 131    &10(1st Aut.)/7(2st. Aut.)          & 5                    \\
ScienceDirect & 36        & 36     &8(1st Aut.)/6(2st Aut.)          & 3                    \\
Scopus        & 258       & 61     &10(1st Aut.)/9(2st Aut.)         & 5                    \\
Total         & 867       & 538    &60/50          & 31 + 4 (Snowballing) \\ \hline
\end{tabular}%
}
\vspace{-1em}
\end{table}

\subsubsection{Searching Results}
The search process was conducted in April 2025, and therefore, the search results reflect studies available up to that date. We present the details of paper counts among different databases in Table~\ref{tab:searching_results}. The total number of papers after filtering and snowballing is 35 (see Appendix~\ref{app: appendixA}).

\section{Classification of PE in RE}

\subsection{Existing Classification in PE}
\label{sec:pe_cate}

\CRV{PE, based on different perspectives, is broadly categorized into different taxonomies~\cite{schulhoff2024prompt,santu2023teler,sasaki2024systematic}. They introduce various types of categorization methods, but presenting these categories at the same hierarchical level can confuse readers: for example, few-shot learning is often used within code generation, yet it appears alongside “code generation” rather than as a subcategory. Here are two representative and typical survey for PE techniques.}
Sahoo et al. \cite{sahoo2024systematic} analyze PE across function, technique, and application, and they propose twelve categories: 1) new tasks without extensive training, 2) reasoning and logic, 3) reduce hallucination, 4) user interaction, 5) fine-tuning and optimization, 6) knowledge-based reasoning and generation, 7) improving consistency and coherence, 8) managing emotions and tone, 9) code generation and execution, 10) optimization and efficiency, 11) understanding user intent, and 12) metacognition and self-reflection. While this taxonomy provides broad coverage, it still mixes different levels of abstraction—“reduce hallucination,” for instance, is a technique that overlaps with “knowledge-based reasoning and generation” as methods to prevent misinformation, making it difficult to distinguish technique from application. From this observation, we conclude that the categorization of PE approaches should follow a clearer logical hierarchy, separating techniques, objectives, and application domains. Fagbohun et al. \cite{fagbohun2024empirical} offer an alternative taxonomy driven by function rather than by individual characteristics or techniques: they define seven functional categories—logical and sequential processing, contextual understanding and memory, specificity and targeting, metacognition and self-reflection, directional and feedback, multimodal and cross-disciplinary, and creative and generative. By emphasizing functional groupings rather than isolated methods like “reduce hallucination,” this framework presents a more standardized, interdisciplinary structure. Nevertheless, its inclusive definitions can also blur boundaries—for example, including code generation under ``metacognition and self-reflection'' lacks a clear conceptual link, since code generation does not inherently involve self-reflective processes. Therefore, even a function-based taxonomy must ensure that each category genuinely encompasses techniques sharing the same underlying function.

\subsection{PE Categorization}

\begin{figure*}[htbp]
    \centering
    \includegraphics[width=0.8\linewidth]{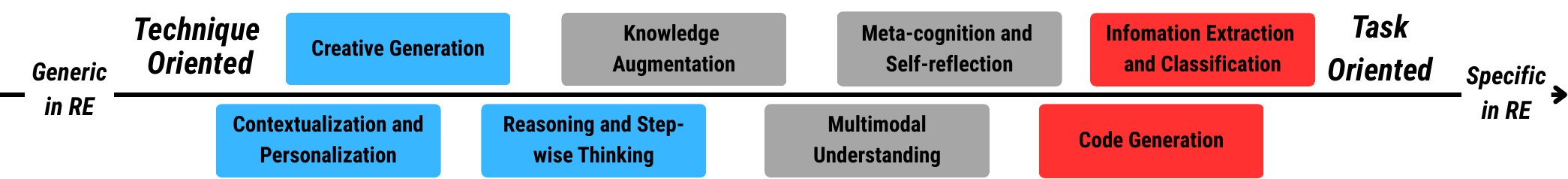}
    \vspace{-0.5em}
    \caption{PE Categorization: Blue are technique-oriented categories, Grey enable both techniques and task-oriented, and Red are task-oriented categories. 
    }
    \label{fig:taxonomy}
    \vspace{-1.5em}
\end{figure*}

\CRV{Inspired by existing PE surveys, we propose a hybrid PE taxonomy (Fig.~\ref{fig:taxonomy}) that spans \textit{task-oriented} to \textit{technique-oriented} approaches by consolidating schemes from Sahoo et al.~\cite{sahoo2024systematic} and Fagbohun et al.~\cite{fagbohun2024empirical}. We transform their “reasoning and logic” into “Reasoning and Stepwise Thinking” and merge overlapping categories (e.g., “reduce hallucination”). Each category is then illustrated using a common smart home thermostat scenario: \textit{In this scenario, you’re designing a smart home thermostat. You should meet the following requirements:}}

$\bullet$ \textbf{\textit{measure ambient temperature,}}

$\bullet$ \textbf{\textit{display it on a screen,}}

$\bullet$ \textbf{\textit{let users set a target temperature,}}

$\bullet$ \textbf{\textit{adjust the HVAC system to reach that target,}}

$\bullet$ \textbf{\textit{log the temperature once per minute, and}}

$\bullet$ \textbf{\textit{send an alert if temperature deviates $>5$ degree from target for $>10$ min.}}

\sectopic{Creative Generation.} The motivation behind creative generation prompting is to push LLMs beyond factual or routine responses, encouraging novelty and innovation in output. In RE, stakeholders are routinely asked to imagine a future that does not yet exist. Creative generation can help produce engaging requirement artifacts, alternative requirement formulations, elicit out-of-the-box feature ideas, and innovative solutions to requirement conflicts. 
In the example below, we show how creative generation can be used in RE. The creative burst can be followed by an evaluation step to establish the feasibility and business value, then translate promising concepts into precise requirements, e.g., ``System shall update LED strip intensity within 500 ms of a detected ±0.5 °C change.''


\begin{lstlisting}[style=prompt,caption={Creative Generation}]
<USER>:
Inspire our design team! Brainstorm an imaginative new feature for a next-gen smart thermostat. Present it as a short, one-sentence bullet point.
<LLM>:
1. Aurora-Glow Mode-paints the room with gentle colour waves that ebb and flow with temperature shifts, turning climate control into ambient art.
2. Whisper-Forecast-murmurs tomorrow's weather in a soft chime each dawn, nudging the thermostat to pre-warm or pre-cool before you even wake.
3. Harmony-Pulse Sync-listens to household music and subtly adjusts airflow rhythm so the room ``breathes'' in time with your favourite playlist.
\end{lstlisting}
\vspace{-0.5em}

\sectopic{Contextualization and Personalization}. This category encompasses influential prompting techniques, including one-shot, few-shot, and in-context learning. These techniques are also useful and commonly adopted in RE~\cite{arora2024generating}. RE involves multiple stakeholders with different backgrounds, concerns, and technical expertise. Each stakeholder group needs requirements presented in their context and terminology. By anchoring generation in this tailor-made context, these prompts markedly improve consistency, reduce redundant explanations, and allow the assistant to adopt a stable voice that adapts to each user's knowledge level and style, thereby boosting engagement and trust in long-running dialogues. Typical techniques include in-context prompts~\cite{rubin2021learning}, multi-persona or role prompts, emotion prompting~\cite{li2023large}, and conversational summarisation buffers, which prepend recent turns, user preferences, or persona instructions before each generation step. 

\begin{lstlisting}[style=prompt,caption={Contextualization and Personalization}, literate={±}{{$\pm$}}1]
<USER>: Example (two-audience list)  
Q: List two key requirements for a humidity sensor for two stakeholder perspectives:
Homeowner (1) and HVAC engineer (2) 
   1. - Warns you when air is too damp or dry  
      - Shows humidity on your phone  
   2. - ±2 % RH accuracy at 25 degree
      - Publishes readings every 60 s via MQTT/TLS  
Task: List two key requirements for a smart thermostat in the same way.
<LLM>:
1. - Keeps your rooms at the temperature you set  
   - Lets you change settings from a mobile app  
2. - Maintains ±0.5 degree via PID control at 1 Hz  
   - Logs data and supports OTA firmware updates
\end{lstlisting}

\sectopic{Reasoning and Step-wise Thinking}. RE sub-tasks often involve complex logical reasoning to derive functional requirements from business needs, trace dependencies between requirements, and validate requirement consistency. LLMs can struggle with multi-step requirement analysis tasks, preferring shortcuts over systematic RE methodologies. To address this issue, Reasoning and Stepwise Thinking PE researchers break the complex reasoning tasks into stepwise reasoning paths to let LLM think through a logical reasoning chain. The typical approach under this category is CoT series~\cite{wei2022chain,zhang2022automatic,zhao2023enhancing}. These CoT-styled prompt templates are widely employed in various tasks, including Tree-of-Thoughts~\cite{yao2023tree}, Graph-of-Thoughts~\cite{yao2023beyond}, and so on. These techniques can be widely employed in different stages of RE, such as requirements decomposition \& validation, dependency analysis, etc.

\begin{lstlisting}[style=prompt,caption={Example: Reasoning and Step-wise Thinking}]
<USER>:
Please identify all functional and non-functional requirements in the thermostat scenario. First, list every candidate requirement. Then, categorize each as FR or NFR. Finally, explaining your reasoning step by step.
<LLM>: 
Step 1: List candidate requirements. Measure ambient temperature - Display temperature on screen - Let user set target temperature - Adjust HVAC to target - Log temperature every minute - Send alert if deviation > 5 degree or > 10 min.
Step 2: Categorize. Measure ambient temperature - FR (it's a core function) - Log temperature every minute - NFR (performance: frequency) - etc.
\end{lstlisting}
\vspace{-0.5em}

\sectopic{Knowledge Augmentation.} RE activities benefit significantly from external knowledge sources, including industry standards, regulatory requirements, best practices, and domain expertise that may not be captured in LLM training data. Knowledge-augmentation prompts. Therefore, inject trustworthy external information—most often via Retrieval-Augmented Generation (RAG)~\cite{lewis2020retrieval} pipelines or tool-use schemas—to ground the model’s reasoning on up-to-date sources. Representative techniques include RAG, ReAct~\cite{yao2023react} and Chain-of-Knowledge prompting \cite{li2023chain}, which prepend retrieved passages, interleave “thought–action–observation” traces, or have the model enumerate supporting facts step-by-step. By explicitly supplying evidence within the prompt, these methods reduce hallucinations, increase factual accuracy, and make answers verifiable, which can be applied in some RE stages that require context information, such as integrating regulatory requirements and domain knowledge. The knowledge augmentation prompt must be based on some external knowledge resource. By performing the information retrieval method, the related context can be used for prompt construction.

\begin{lstlisting}[style=prompt,caption={Example: Knowledge Augmentation}]
<CONTEXT>:
Assume the extracted content from the external knowledge resource (ASHRAE Handbook):
    Best practice: log temperature data at 5 min intervals for energy audits; alarms recommended at 4 degree deviation over 15 min.
<USER>:
Using the ASHRAE excerpt above, update the thermostat's NFRs (logging interval and alert thresholds) and justify each change.
<LLM>:
Logging Interval: Change from 1 min to 5 min. Justification: Aligns with ASHRAE energy audit standards to reduce data storage overhead while still capturing enough resolution.
Alert Threshold: Change from 5 degree over 10 min to 4 degree over 15 min. Justification: Matches industry guidelines for fault detection without triggering nuisance alarms during normal HVAC cycling.
\end{lstlisting}
\vspace{-0.5em}

\sectopic{Multimodal Understanding.} RE widely accepts various types of specifications, including UI mockups, system diagrams, process flows, and prototype demonstrations that need to be analyzed alongside textual requirements. Prevailing frameworks either (i) pair specialized vision or audio encoders with a language decoder via a lightweight bridging module, as in BLIP-2~\cite{li2023blip}, or (ii) train a single Transformer end-to-end on interleaved modality tokens, exemplified by Flamingo~\cite{alayrac2022flamingo}, Kosmos-1~\cite{huang2023language} \& Kosmos-2~\cite{peng2023kosmos}, and GPT-4V~\cite{wu2023gpt4vis}. While offering greater robustness and context fidelity, multimodal models incur higher computational cost, risk modality imbalance, and can compound data-borne biases. Overall, even multi-modal prompting techniques are not commonly adopted in the RE community; however, they can still be applied in most stages, particularly for requirements elicitation from domain documents and meta-model-driven software design.

\begin{lstlisting}[style=prompt,caption={Example: Multimodal Understanding}]
<FIGURE>:
(An image will be attached in the prompt, which is a screenshot of the thermostat's touchscreen UI.)
<USER>:
Here's the current UI mockup. Identify three usability issues that might confuse users, then suggest a quick redesign note for each.
<LLM>:
Tiny Font for Temperature: Difficult to read at a glance. Redesign: Enlarge digits to at least 24 px and use high-contrast colors.
Unlabeled Buttons: ``+'' and ``-'' icons without context. Redesign: Add tooltip or label (``Set Temp +/-'').
No Status Indicator for Logging: Users can't tell if logging is active. Redesign: Include a small ``Logging'' LED icon in the corner.
\end{lstlisting}
\vspace{-0.5em}

\sectopic{Meta-cognition and Self-reflection.} While LLMs respond to input queries, models may commit logical slips or confidently output flawed facts. Self-reflective prompting forces them to critique and, if needed, revise those first drafts, which can assure the quality of the response. Quality is also critical in RE, as requirement defects are expensive to fix later in development. Self-reflection helps ensure requirement completeness, consistency, and quality through systematic review processes. Core approaches—such as Reflexion~\cite{shinn2023reflexion}, Meta-Cognition~\cite{wang2023metacognitive}, Self-Explanation~\cite{gao2023self} and Step-Back prompting~\cite{zheng2023take}, either generate multiple independent solutions and elect the majority answer, or insert an explicit loop from critique to revise where the model reviews its own reasoning trace for gaps, bias or factual errors before finalising. The most typical application scenario can be requirements review. The key feature of this category is that the prompt will utilise multiple rounds of instruction to allow LLMs to reconsider and double-check their responses.

\begin{lstlisting}[style=prompt,caption={Example: Meta-cognition and Self-reflection}]
<PREVIOUS CONVERSATION>:
(First Round) First, we ask it to extract FRs/NFRs (as in the example of Reasoning and Step-wise Thinking.)
<USER>:
You just pulled out six requirements. Are you confident none are missing? If any gaps exist, critique and add them.
<LLM>:
On reflection, I omitted the security requirement. FR: Authenticate user before allowing target-temp changes. Rationale: Prevents unauthorized HVAC control. Now there are seven requirements in total.
\end{lstlisting}
\vspace{-0.5em}

\sectopic{Information Extraction and Classification.}
RE practitioners sometimes need to extract, categorize, and structure requirements from various sources, including stakeholder interviews, existing documentation, legacy systems, and regulatory documents. The information extraction and classification refer to extracting or analyzing the external resources, which always cooperate with contextualization and personalization prompting methods in the RAG~\cite{lewis2020retrieval} framework. The application of Information Extraction and Classification in the RE process includes requirements classification and priority classification.

\begin{lstlisting}[style=prompt,caption={Example: Information Extraction and Classification}]
<USER>:
From the scenario text, extract every requirement into CSV with columns ID, Type, Text, where Type is FR or NFR.
<LLM>:
*R1, FR, Measure ambient temperature
*R2, FR, Display temperature on screen
\end{lstlisting}
\vspace{-0.5em}

\sectopic{Code Generation.} Code here refers to a concrete SE artifact, encompassing system implementations, software design, and testing. In many RE projects we need a quick, runnable artefact to check that a requirement
has been understood correctly, before investing weeks in full implementation.
Code generation lets analysts move straight from text to artifacts of interest, closing the gap between ``specify'' and ``validate''. Code demands strict syntax and logical correctness, so prompts must be specially designed. Along with the other prompting techniques, the code generation prompt has two special steps, including (1) context provisioning, which gives the description to LLMs to specify the objectives, and (2) structured specification, which regulates LLMs for the output format, including the scale and the programming language.
For the example below, with a single prompt below we now have an executable specification that stakeholders can run, developers can extend, and auditors can trace—illustrating
the practical value of code-generation prompts in RE.

\begin{lstlisting}[style=prompt,
  caption={Code Generation}, literate={±}{{$\pm$}}1
           {≠}{{$\neq$}}1]
<CONTEXT>
R1, FR, "System shall log temperature every 60 s"
R2, NFR, "Temperature accuracy shall be within 0.5 degree"
<USER>
Write a PyTest file that:
1. Creates one test per requirement.
2. Tags each test with @pytest.mark.req("<ReqID>").
3. Fails if logging interval ≠ 60 s or accuracy drift > 0.5 degree.
<LLM>
# test_thermostat_reqs.py
import thermostat, pytest
@pytest.mark.req("R1")
def test_log_interval():
    assert thermostat.LOG_INTERVAL == 60
@pytest.mark.req("R2")
def test_temp_accuracy():
    diff = abs(thermostat.read_temp() - thermostat.true_temp())
    assert diff <= 0.5
\end{lstlisting}
\vspace{-0.5em}

Our categorization is designed to be mutually exclusive yet collectively exhaustive. By anchoring each category in a distinct primary function, this taxonomy retains the strengths of earlier work while offering sharper conceptual boundaries and providing a clearer map from real‑world prompting goals to the design strategies best suited to achieve them.

\section{Results and Discussion}
\label{sec:results}
\subsection{RQ1 Results}
\sectopic{RQ1-1. What is the end-goal of the selected studies?}
\begin{table}[]
\scriptsize
\centering
\caption{Overview of End-Goals in Selected Studies}
\vspace{-0.5em}
\label{tab:rq1-task}
\begin{tabularx}{\columnwidth}{lXl}
\hline
\textbf{PID} & \textbf{Target Task}                                                        & \textbf{Task Categorization}                  \\ \hline
\textbf{P1}       & Goal Model Extraction                                              & Model Gen.                     \\
\textbf{P2}       & Software Conflict Detection                                        & Software Analysis                    \\
\textbf{P3}       & Software Safety Analysis                                           & Safety Case Gen. and Analysis  \\
\textbf{P4}       & Use Case Classification                                            & Req. Analysis and Gen. \\
\textbf{P5}       & Automated Test Gen.                                          & Test Gen.                      \\
\textbf{P6}      & Software Development ChatBot                                       & SDLC Assistant                       \\
\textbf{P7}       & Req. driven Testing                                        & Test Gen.                   \\
\textbf{P8}       & Eval. of Req. Analysis Against Human                  & Req. Analysis and Gen. \\
\textbf{P9}       & Assurance Case Defeater Identification                           & Defeater Identification              \\
\textbf{P10}      & Enhancement of Code Gen. Capability                          & Code Analysis and Gen.         \\
\textbf{P11}      & Defect Correction in Req.                                  & Defect Correction                    \\
\textbf{P12}      & Domain Model Extraction                              & Model Gen.                     \\
\textbf{P13}      & Enhancement of Use Case                                            & Req. Analysis and Gen. \\
\textbf{P14}      & Enhancement of Use Case Gen.                                 & Req. Analysis and Gen. \\
\textbf{P15}      & Domain Model Extraction                          & Model Gen.                     \\
\textbf{P16}      & Multi-agent Sys. for Elicitation/Analysis       & Req. Analysis and Gen. \\
\textbf{P17}     & RE-driven Java Code Gen.                                               & Code Analysis and Gen.         \\
\textbf{P18}      & Generating Req. Elicitation Interview Scripts              & Req. Analysis and Gen. \\
\textbf{P19}      & Transforming Software Req. into User Stories               & Req. Analysis and Gen. \\
\textbf{P20}      & Sensor Input Setup Assistant                                       & Software Analysis                    \\
\textbf{P21}      & Software Development Assistant for Communication System            & SDLC Assistant                       \\
\textbf{P22}      & RE Assistant for Digital Twins               & Req. Analysis and Gen. \\
\textbf{P23}      & Evolving Req. Elicitation from App Reviews                 & Req. Analysis and Gen. \\
\textbf{P24}      & Natural Language Inference (NLI) in Req. Engineering Tasks & Req. Analysis and Gen. \\
\textbf{P25}      & LLMs for Software Engineering Education and Training               & Software Analysis                    \\
\textbf{P26}      & Req. driven UML Sequence Diagrams                          & Model Gen.                     \\
\textbf{P27}      & Automatic Gen. of Formal Req. Specifications         & Req. Analysis and Gen. \\
\textbf{P28}      & GPT-4 for Creating Goal Models                                     & Model Gen.                     \\
\textbf{P29}      & Automatic Safety Case Gen.                                   & Safety Case Gen. and Analysis  \\
\textbf{P30}      & Req. Modeling Aided by ChatGPT                             & Req. Analysis and Gen. \\
\textbf{P31}      & Req. Satisfiability Analysis                               & Req. Analysis and Gen. \\
\textbf{P32}      & Req.-driven Slicing of Simulink Models                     & Model Gen.                     \\
\textbf{P33}      & Acceptance Case Gen.                                         & Test Gen.                      \\
\textbf{P34}      & Req. Coverage Reviewing                                    & Req. Analysis and Gen. \\
\textbf{P35}      & Automated User Story Gen.                                    & Req. Analysis and Gen. \\ \hline
\end{tabularx}%
\vspace{-1em}
\end{table}
In the selected studies, we identified that there are various types of end-goals and tasks included, i.e., P1 and P26 are both focused on model generation, P10 and P17 are both for code generation. Thus, we present a classification of all the end goals in Table.~\ref{tab:rq1-task} and \ref{tab:rq1-task_cate}.

\noindent $\bullet$ Requirements Analysis and Generation is the most common task in the selected papers (15 studies). These studies directly employ PE to optimize the RE process, including requirements optimization (P13, P14, P22), use case classification (P4), evaluation of requirements analysis (P8, P24, P31, P34), requirements generation (P16, P18, P19, P23, P27, P30, P35). Considering the scope is PE4RE, it is reasonable there are studies directly related to RE. Besides, with the adoption of natural language requirements, such as user stories and use cases,  PE can enhance the processing capabilities of NL by leveraging advancements of LLMs.

\noindent $\bullet$ Model Generation is the second most popular goal in the selected papers (6 studies). In contrast, meta-model requirements specification is widely adopted in the RE process. However, the formulation of meta-model-based requirements is always a challenge in RE domain~\cite{Ehikioya2020ACA, Giraldo2016EvidencesOT}. With the help of the PE methods, the generalization ability of LLMs is exhausted in the model generation task.

\noindent $\bullet$ There are four studies related to the Software Analysis topic. These studies not only focus on the RE field, but also use requirements to assist the analysis of software artefacts, such as conflict detection (P2) and sensor input setup (P20), which have broader application scenarios.

\noindent $\bullet$ Code Analysis and Generation is also an attractive task recently. As a highly structured language, a high-quality prompt template can significantly enhance the code generation ability by minimizing ambiguous input and preventing hallucination.

\begin{table}[]
\scriptsize
\centering
\caption{Categories of Prompting Methods}
\vspace{-0.5em}
\label{tab:rq1-task_cate}
\begin{tabularx}{\columnwidth}{lXc}
\hline
\textbf{Task Categories}                      & \textbf{PIDs}                                                                                               & \textbf{Num.} \\ \hline
\textbf{Model Generation}                     & P1, P12, P15, P26, P28, P32                                                                         & 6    \\
\textbf{Software Analysis}                    & P2, P20, P21, P25                                                                                   & 4    \\
\textbf{Safety Ana.\ \& Gen.}                 & P3, P29                                                                                            & 2    \\
\textbf{Req Ana.\ \& Gen.}                    & P4, P8, P13, P14, P16, P18, P19, P22, P23, P24, P27, P30, P31, P34, P35                            & 15   \\
\textbf{Code Ana.\ \& Gen.}                   & P10, P17                                                                                           & 2    \\
\textbf{Test Generation}                      & P5, P7, P33                                                                                        & 3    \\
\textbf{SDLC Assistant}                       & P6, P21                                                                                            & 2    \\
\textbf{Defeater Identification}              & P9                                                                                                 & 1    \\
\textbf{Defect Correction}                    & P11                                                                                                & 1    \\ \hline
\end{tabularx}%
\end{table}

\sectopic{RQ1-2. Which LLMs are adopted in the selected papers?} 
\begin{table}[]
\scriptsize
\centering
\caption{The Number of LLMs Adopted in Selected Studies}
\vspace{-0.5em}
\label{tab:rq1-LLM}
\begin{tabularx}{\columnwidth}{lXc}
\hline
\textbf{LLM Types} & \textbf{PIDs}                                                                                                                                                                                                                                                         & \textbf{Num.} \\ \hline
\textbf{GPTs}          & P1 (GPT4), P3 (GPT4o), P4 (GPT3.5-Turbo), P5 (GPT4), P6 (GPT3.5-Turbo, GPT4), P7 (GPT3.5-Turbo, GPT4), P8 (GPT3.5-Turbo, GPT4), P9 (GPT3.5-Turbo), P10 (GPT3.5-Turbo, GPT4), P11 (GPT3.5, GPT4), P12 (GPT4), P13 (GPT3.5-Turbo), P14 (GPT4-Turbo, GPT3.5-Turbo, GPT3.5-Turbo-instruction), P15 (GPT3.5-Turbo), P16 (GPT3.5, GPT4), P17 (GPT3.5-Turbo), P18 (GPT3.5), P19 (GPT3.5), P20 (GPT4o), P21 (GPT3.5), P22 (GPT3.5), P23 (GPT4), P24 (GPT3.5, GPT4), P25 (GPT3.5), P26 (GPT3.5), P27 (GPT3.5), P28 (GPT4), P29 (GPT4), P30 (GPT3.5-Turbo, GPT4), P31 (GPT3.5-Turbo, GPT4), P32 (GPT3.5-Turbo), P33 (GPT3.5-Turbo), P34 (GPT3.5, GPT4), P35 (GPT4-Turbo) & 34            \\
\textbf{Llama}         & P1 (Llama-13B), P4 (Llama2-70B), P5 (Llama2-70B), P10 (CodeLlama-34B), P25 (LLaMA-2-70B), P33 (Llama-7B \& 13B)                                                                                                                                               & 6             \\
\textbf{Other}         & P1 (Cohere-Command), P8 (Nous-Hermes-2), P10 (CodeGeeX), P11 (ERNIE-Bot), P14 (Whisper), P18 (Bard), P27 (T5-Large), P33 (text-davinci-003)                                                                                                                         & 8             \\
\textbf{PLM}           & P2 (BERT, ALBERT, RoBERTa, XLNet), P11 (BERT), P23 (BERT), P24 (RoBERTa, NoRBERT, BERT)                                                                                                                                                                             & 4             \\
\textbf{Qwen}          & P3 (Qwen-Max)                                                                                                                                                                                                                                                   & 1             \\
\textbf{Gemini}        & P5 (Gemini-Pro)                                                                                                                                                                                                                                                  & 1             \\
\textbf{Mistral}       & P5 (Mistral-7B), P25 (Mixtral-8×7B)                                                                                                                                                                                                                               & 2             \\
\textbf{Phi}           & P8 (Phi-3-Medium)                                                                                                                                                                                                                                                & 1             \\
\textbf{Deepseek}      & P10 (DeepSeek-Coder-V2)                                                                                                                                                                                                                                           & 1             \\
\textbf{Claude}        & P21 (Claude 3)                                                                                                                                                                                                                                                   & 1             \\ \hline
\end{tabularx}%
\vspace{-1em}
\end{table}
In this sub-RQ, we present all the applied LLMs in the selected papers. Due to different preferred input style among various LLMs, the selection of LLMs in PE research can somehow reflect the generalization ability of prompting methods. We present a comprehensive list of adopted LLMs in Table.~\ref{tab:rq1-LLM}, which suggests that,

\noindent $\bullet$ GPT-series is the dominant choice in the selected papers. As the most popular LLM, GPTs provide one of the friendliest interaction websites and interfaces, stable API inference, and various plugins. The common preference for GPTs reflects their strong generalization ability and their representative roles among all LLMs. Since prompting is typically tied to training or fine-tuning, the PE method will remain effective even if GPTs do not support gradient visibility.

\noindent $\bullet$ Llama is in second place in LLM selection. Different from GPTs, Llama is an open-source LLM that supports customization and fine-tuning.

\noindent $\bullet$ Some task-specific LLMs are adopted in the selected paper, i.e., Copilot, CodeLlama, and DeepSeek-Coder, which are known for their leading role in code generation area, the selection of these LLMs aligns with the end-task in RQ1-1, which suggests that prompt designing should widely consider both task-specific and generic LLMs.

\sectopic{RQ1-3. How are PE4RE studies in the context of PE categorization?}
As we introduced our categorization method in Section\ref{sec:pe_cate}, we further map the selected studies into our defined categorizations, which are shown in Table\ref {tab:rq1-prompt} and \ref{tab:rq1-prompt_cate}.
Based on the results, we found that Contextualisation \& Personalisation (26 studies) and Reasoning \& Step-wise Thinking (20 studies) are the top-2 adopted prompt-engineering methods in the selected studies, followed by Creative Generation (14 studies), Information Extraction \& Classification (11 studies), Knowledge Augmentation (8 studies), Meta-cognition \& Self-reflection (6 studies), Code Generation (4 studies), Defeater Identification / Defect Correction (1 studies), Multimodal-Understanding (0 studies).
As for the co-occurrence patterns between the multiple types of prompt-engineering methods, we have the following conclusions, (1) Contextualisation \& Personalisation and Reasoning \& Step-wise Thinking are always adopted together (14/20), which suggest that the dominant recipe is role prompt + chain-of-thought, suggesting researchers view rich context as a prerequisite for trustworthy reasoning; (2) Knowledge Augmentation methods always adopted with the other techniques, as 4/8 RAG-style papers also include Reasoning prompts, which indicates that researchers tend to use retrieved snippets as extra context rather than as a distinct prompting paradigm; (3) Meta-cognition \& Self-reflection methods are almost always layered on top of Reasoning \& Step-wise Thinking, where verifier–reviser or self-consistency loops require step-wise traces to critique, reinforcing the centrality of explicit reasoning in RE prompts.
\begin{table}[]
\centering
\scriptsize
\caption{Overview of Prompting Methods}
\vspace{-0.5em}
\label{tab:rq1-prompt}
\begin{tabularx}{\columnwidth}{lX}
\hline
\textbf{PID} & \textbf{Prompting Methods} \\ \hline
\textbf{P1}  & \textbf{Reasoning \& Step-wise Thinking} (chain-of-thought style 7-step prompts)\textbf{; Contextualization \& Personalization} (role-prompt ``You are a RE engineer…'', domain background supplied) \\
\textbf{P2}  & \textbf{Info. Extraction \& Classification} (classifies SRS pairs into Conflict / Duplicate / Neutral) \\
\textbf{P3}  & \textbf{Knowledge Augmentation} (retrieval-augmented Gen. against safety regulations)\textbf{; Meta-cognition \& Self-reflection} (Researcher \& Revisor loop)\textbf{; Contextualization \& Personalization} (agents given explicit functional-safety roles) \\
\textbf{P4}  & \textbf{Info. Extraction \& Classification} \\
\textbf{P5}  & \textbf{Reasoning \& Step-wise Thinking} (CoT)\textbf{; Contextualization \& Personalization} (few-shot examples) \\
\textbf{P6}  & \textbf{Reasoning \& Step-wise Thinking} (ReAct with explicit action steps)\textbf{; Contextualization \& Personalization} (conversation history + module state) \\
\textbf{P7}  & \textbf{Knowledge Augmentation} (RAG pipeline)\textbf{; Contextualization \& Personalization} (optional exemplar scenario = few-shot) \\
\textbf{P8}  & \textbf{Contextualization \& Personalization} (few-shot practice doc + example defects) \\
\textbf{P9}  & \textbf{Contextualization \& Personalization} (role-based system prompt) \\
\textbf{P10} & \textbf{Reasoning \& Step-wise} (Structured CoT)\textbf{; Contextualization} (role-definition prompts)\textbf{; Meta-cognition \& Self-reflection} (internal code-review stage)\textbf{; Code Gen.} \\
\textbf{P11} & \textbf{Contextualization \& Personalization} (rich upstream cues)\textbf{; Info. Extraction \& Classification} (defect detection)\textbf{; Knowledge Augmentation} (domain-specific keywords)\textbf{; Creative Gen.} (rewriting correct sentences) \\
\textbf{P12} & \textbf{Info. Extraction \& Classification}\textbf{; Reasoning \& Step-wise Thinking} (OpenAI and existed prompt patterns)\textbf{; Contextualization \& Personalization} (guidance examples) \\
\textbf{P13} & \textbf{Contextualization \& Personalization} (persona pattern)\textbf{; Meta-cognition \& Self-reflection} (feedback loop \& self-revision)\textbf{; Creative Gen.} (rewriting use-cases) \\
\textbf{P14} & \textbf{Creative Gen.}\textbf{; Contextualization \& Personalization} \\
\textbf{P15} & \textbf{Info. Extraction \& Classification}\textbf{; Reasoning \& Step-wise Thinking} \\
\textbf{P16} & \textbf{Contextualization \& Personalization} (role prompts)\textbf{; Reasoning \& Step-wise Thinking} \\
\textbf{P17} & \textbf{Code Gen.}\textbf{; Knowledge Augmentation}\textbf{; Contextualization \& Personalization}\textbf{; Reasoning \& Step-wise Thinking} \\
\textbf{P18} & \textbf{Reasoning \& Step-wise Thinking}\textbf{; Creative Gen.}\textbf{; Contextualization \& Personalization} \\
\textbf{P19} & \textbf{Creative Gen.}\textbf{; Contextualization \& Personalization} (few-shot examples) \\
\textbf{P20} & \textbf{Knowledge Augmentation}\textbf{; Contextualization \& Personalization}\textbf{; Reasoning \& Step-wise Thinking}\textbf{; Info. Extraction \& Classification} \\
\textbf{P21} & \textbf{Code Gen.}\textbf{; Reasoning \& Step-wise Thinking}\textbf{; Contextualization \& Personalization}\textbf{; Meta-cognition \& Self-reflection} (inspector-critique loop) \\
\textbf{P22} & \textbf{Creative Gen.} (LLM-authored requirements)\textbf{; Reasoning \& Step-wise Thinking} (iterative prompting to improve completeness \& correctness) \\
\textbf{P23} & \textbf{Contextualization \& Personalization} (R-T-O-QA template grounds the task)\textbf{; Knowledge Augmentation} (300 k-entry sentiment lexicon injected)\textbf{; Info. Extraction \& Classification} (aspect-based SA) \\
\textbf{P24} & \textbf{Info. Extraction \& Classification} (entailment)\textbf{; Reasoning \& Step-wise Thinking} (logic of entail \& contradict)\textbf{; Contextualization \& Personalization} (label verbalisation = rich hypotheses) \\
\textbf{P25} & \textbf{Creative Gen.} (idea lists, code)\textbf{; Contextualization \& Personalization} (role-play, tutor personas)\textbf{; Reasoning \& Step-wise Thinking} (effort-estimation dialogue) \\
\textbf{P26} & \textbf{Creative Gen.} (diagram synthesis)\textbf{; Reasoning \& Step-wise Thinking} (LLM must infer interactions) \\
\textbf{P27} & \textbf{Creative Gen.} (dataset Gen.)\textbf{; Info. Extraction \& Classification} (AP recognition)\textbf{; Code Gen.} (CTL code output) \\
\textbf{P28} & \textbf{Creative Gen.} (model Gen.)\textbf{; Contextualization \& Personalization} (injecting domain context) \\
\textbf{P29} & \textbf{Creative Gen.}\textbf{; Contextualization \& Personalization}\textbf{; Info. Extraction \& Classification} \\
\textbf{P30} & \textbf{Info. Extraction \& Classification} (entity \& relation extraction)\textbf{; Creative Gen.} (diagram synthesis)\textbf{; Contextualization \& Personalization} (prompt embeds element definitions) \\
\textbf{P31} & \textbf{Reasoning \& Step-wise Thinking} (chain-of-thought)\textbf{; Knowledge Augmentation} (knowledge snippets)\textbf{; Creative Gen.} (specification Gen.)\textbf{; Contextualization \& Personalization} (requirement inversion) \\
\textbf{P32} & \textbf{Reasoning \& Step-wise Thinking} (CoT)\textbf{; Knowledge Augmentation} (model text injected) \\
\textbf{P33} & \textbf{Reasoning \& Step-wise Thinking} (K-shot)\textbf{; Contextualization \& Personalization} (role persona) \\
\textbf{P34} & \textbf{Reasoning \& Step-wise Thinking} (CoT)\textbf{; Meta-cognition \& Self-reflection} (ask model to justify in 10 words) \\
\textbf{P35} & \textbf{Reasoning \& Step-wise Thinking} (CoT)\textbf{; Meta-cognition \& Self-reflection} (refine phase)\textbf{; Creative Gen.} (new user stories) \\ \hline
\end{tabularx}%
\vspace{-1em}
\end{table}

\subsection{RQ2 Results}

\sectopic{RQ2-1: What are the limitations and challenges in current PE4RE studies?} In this sub-RQ, we aim to identify the limitations and challenges in our selected studies and the potential influence of these challenges. Across the 35 primary studies we analysed, we identified limitations as,

\noindent $\bullet$ \textbf{L1. No studies employ non-textual prompting methods}, such as Multi-Modal Understanding, to process non-textual input. We notice that, except for P26 and P30, which produce UML, P32 converts models to text, the other studies predominantly concentrate on code-based or textual tasks. As an important part of RE, model-based representation is not frequently explored in PE4RE research, which should be investigated in the future.

\noindent $\bullet$ \textbf{L2. Lack of investigation in requirements elicitation}. We found that in the selected studies, none of the papers formally report requirements generation or elicitation. Only four papers (P15, P27, P25, P33) prompt LLMs to ask questions, conduct interviews, or simulate stakeholder interactions. The remaining studies primarily focused on later stages, assuming that requirements existed.
    
\noindent $\bullet$ \textbf{L3. Lack of Systematic Evaluation}. Only 2 (P1, P31) of the 20 Reasoning \& Step-wise papers conducted ablation studies. As for Knowledge Augmentation, only eight papers use RAG. In P3 and P20, adding one retrieved chunk improves coverage and SUS scores; however, P7 shows that injecting more than one chunk lowers BLEU/ROUGE. We think the evaluation of the PE4RE paper cannot show a standard evaluation process.

\begin{tcolorbox}[colback=green!10!,
    colframe=white, breakable]
\noindent \textbf{Maturity gap}: While advanced NLP research explores retrieval, self-reflection, vision–language fusion, and cost governance, most PE4RE work still centres on single-prompt templates evaluated on edge cases. The result is promising prototypes that remain difficult to replicate, audit, or certify in industrial RE workflows.
\end{tcolorbox}

\begin{table}[]
\scriptsize
\centering
\caption{Categorization of Prompting Methods in Selected Studies}
\vspace{-0.5em}
\label{tab:rq1-prompt_cate}
\begin{tabularx}{\columnwidth}{lXc}
\hline
\textbf{PE Categories}                            & \textbf{PIDs}                                                                                                                           & \multicolumn{1}{l}{\textbf{Num.}} \\ \hline
\textbf{Reasoning \& Step-wise Thinking}          & P1, P5, P6, P10, P12, P15, P16, P17, P18, P20, P21, P22, P24, P25, P26, P31, P32, P33, P34, P35              & 20                       \\
\textbf{Contextualization \& Personalization}       & P1, P3, P5, P6, P7, P8, P9, P10, P11, P12, P13, P14, P16, P17, P18, P19, P20, P21, P23, P24, P25, P28, P29, P30, P31, P33 & 26                       \\
\textbf{Info. Extraction \& Classification} & P2, P4, P11, P12, P15, P20, P23, P24, P27, P29, P30                                                                        & 11                       \\
\textbf{Knowledge Augmentation}                   & P3, P7, P11, P17, P20, P23, P31, P32                                                                                            & 8                        \\
\textbf{Meta-cognition \& Self-reflection}        & P3, P10, P13, P21, P34, P35                                                                                            & 6                        \\
\textbf{Code Generation}                          & P10, P17, P21, P27                                                                                           & 4                        \\
\textbf{Creative Generation}                      & P11, P13, P14, P18, P19, P22, P25, P26, P27, P28, P29, P30, P31, P35                                                            & 14                       \\
\textbf{Multimodal Understanding}                 &                                                                                                                                 & 0                        \\
\textbf{Defeater Identification}                  & P9                                                                                                                            & 1                        \\
\textbf{Defect Correction}                        & P11                                                                                                                            & 1                        \\ \hline
\end{tabularx}%
\vspace{-1em}
\end{table}

\sectopic{RQ2-2. Road-map and Practical Guidelines for PE4RE.}
Building on the gaps highlighted in RQ2-1, we propose a four-stage roadmap that
can progressively advance PE4RE from
prototype studies to an industry-ready discipline.

\noindent $\bullet$ \textbf{R1. Multimodal Elicitation \& Consistency Checking}  
      \textit{(addresses limitation L1: ``text-only focus'')}  
Tools such as GPT-4o, Gemini 2.5 Pro, and LLaVA-Next now accept images and structured graphics. A promising research direction is to combine UML, BPMN, or SysML diagrams with NL requirements, enabling the model to identify inconsistencies or generate trace links without losing any information.

      \begin{itemize}
        \item \emph{Research task}: Fuse textual prompts with visual artefacts, e.g.\ UML, wire-frames, or SysML block diagrams to enable cross-modal traceability and
              inconsistency detection.  
        \item \emph{Starter experiment}: Re-implement P30’s sequence-diagram generator
              using GPT-4o or Gemini 2.5 Pro with native SVG/XMI input; measure precision
              and recall of link predictions against a gold standard.  
        \item \emph{Deliverable}: A public benchmark of 100+ multimodal RE cases with labelled links (ground truth).
      \end{itemize}

\noindent $\bullet$ \textbf{R2. Elicitation-Phase Prompt Patterns} 
      \textit{(addresses limitation L2: “elicitation under-explored”)}  Requirements elicitation is a crucial stage in the RE process. Although this stage is not thoroughly addressed in LLM-based automation, some RAG methods or knowledge augmentation methods can significantly facilitate the requirements generation process.
      \begin{itemize}
        \item Combine \emph{agent templates} (P16) with multi-step \emph{reasoning cues} (P1) to create conversational agents that interview stakeholders, negotiate priorities, and output user stories.  
        \item Evaluate with think-aloud studies: 10 analysts × 3 projects; measure time-to-first-draft, and requirement completeness level $F_{1}$.  
        \item Curate a library of reusable prompts (e.g.\ NFR checklist,
              risk-oriented probes) to standardise elicitation quality.
      \end{itemize}

\noindent $\bullet$ \textbf{R3. End-to-End Traceability and Impact Analysis}  
      \textit{(addresses limitation L3: “lack of systematic evaluation”)}
      
      \begin{itemize}
        \item Retrofit retrieval pipelines (P3, P20) so that each prompt embeds a unique
              requirement ID; any model-generated change triggers an automatic “blast-radius” report identifying affected tests, designs, and code.  
        \item Define an evaluation protocol: given $N$ changed requirements, measure
              (i) recall of impacted artefacts, (ii) false-positive rate, and
              (iii) review effort (min).  
        \item Release evaluation scripts and Docker images to encourage replication
              studies.
      \end{itemize}

\noindent $\bullet$ \textbf{R4. Community Benchmark Suite \& Reporting Checklist}  
      \textit{(cross-cuts all limitations)}  
      
      \begin{itemize}
        \item Aggregate datasets from R1–R3 plus existing corpora (P1, P26, P30) into a single benchmark covering \emph{text}, \emph{code}, and \emph{diagram}
              tasks.  
        \item Publish a PE4RE reporting checklist (e.g., model, prompt template, context length, model params, and metrics).  
        \item Host an annual shared task (e.g.\ at RE ’26) to drive comparable results and foster tooling ecosystems.
      \end{itemize}

\begin{tcolorbox}[colback=green!10!,
    colframe=white, breakable]
    \noindent\textbf{Expected Pay-offs.}  
R1–R4 create a virtuous cycle: richer multimodal LLMs (R1) enable smarter
elicitation agents (R2); explicit IDs and retrieval (R3) support reliable,
auditable pipelines, while a benchmark suite (R4) grounds future work in rigorous,
replicable evaluation. Collectively, these milestones move PE4RE from
isolated proofs-of-concept to a \emph{standardised, stakeholder-centred, and
evidence-based RE practice}.
\end{tcolorbox}

\section{Threats to Validity}
\textbf{\textit{Internal Validity.}} The first author designed the SLR protocol, which was collaboratively reviewed and refined by all co-authors. Search strings were iteratively adjusted and customized for each database to optimize retrieval. Study selection followed a three-stage filtering process: (1) title and abstract screening, (2) brief reading with keyword matching, and (3) comprehensive paper review. All authors validated the final selection to ensure robustness. Data extraction was conducted using Google Forms, with all authors participating in pilot testing to ensure consistency. \textbf{\textit{Construct Validity.}} We mitigated threats through automated and manual searches across six scientific databases. All authors refined inclusion/exclusion criteria through extensive discussions. The first and second authors carefully reviewed studies with vague methodologies to reach consensus on inclusion. \textbf{\textit{Conclusion Validity.}} Threats were minimized through validated search and data extraction processes. The data extraction form was designed based on predefined RQs and verified by co-authors before full implementation. Multiple discussions during analysis ensured robust categorization and meaningful conclusions. \textbf{\textit{External Validity.}} We employed comprehensive search strategies following established guidelines. Our focus on peer-reviewed English studies excluded grey literature and opinion pieces to minimize bias, though this may limit some relevant works. 

\section{Conclusion}
\label{sec:conclusion}
This work presents the first systematic review of Prompt Engineering for Requirements Engineering (PE4RE), analyzing 35 primary studies from 2018-2025. We propose novel categorizations of PE techniques from technique-oriented and task-oriented perspectives, revealing current trends in target tasks, LLMs employed, and prompting methods.
Our analysis identifies critical research gaps and establishes a roadmap for future PE4RE development. Our guidelines aim to advance PE4RE from random prototypes to comprehensive, industry-standard solutions by consolidating current practices and outlining future directions. We anticipate follow-up work to empirically validate these recommendations, create open benchmark datasets, and incorporate PE4RE tools into mainstream RE toolchains. This approach will help accelerate the adoption of LLM-driven methods in real-world SE scenarios.

\bibliographystyle{IEEEtran}
\bibliography{references}
\appendix
\section{Appendix A: Included Primary Studies}
\label{app: appendixA}
\scriptsize
\noindent
\begin{prlist}
  \item Siddeshwar, Vaishali; Alwidian, Sanaa; Makrehchi, Masoud.  
        A Comparative Study of Large Language Models for Goal Model Extraction.  
        In \textit{Proceedings of the ACM/IEEE 27th International Conference on Model Driven Engineering Languages and Systems}, ACM, 2024, pp.\,253–263.

  \item Helmeczi, Robert K.; Cevik, Mucahit; Yıldırım, Savas.  
        A Prompt-based Few-shot Learning Approach to Software Conflict Detection.  
        In \textit{Proceedings of the 32nd Annual International Conference on Computer Science and Software Engineering}, IBM Corp., 2022, pp.\,101–109.

  \item Shi, Lu; Qi, Bin; Luo, Jiarui; Zhang, Yang; Liang, Zhanzhao; Gao, Zhaowei; Deng, Wenke; Sun, Lin.  
        Aegis: An Advanced LLM-Based Multi-Agent for Intelligent Functional Safety Engineering.  
        In \textit{EMNLP 2024 – Conference on Empirical Methods in Natural Language Processing (Industry Track)}, 2024, pp.\,1571–1583.

  \item Alawaji, Batool; Hakami, Mona; Alshemaimri, Bader.  
        Evaluating Generative Language Models with Prompt Engineering for Categorizing User Stories to Its Sector Domains.  
        In \textit{2024 IEEE 9th International Conference for Convergence in Technology (I2CT)}, IEEE, 2024, pp.\,1–8.

  \item S, Shakthi; Srivastava, Pratibha; L, Ravi Kumar; Prasad, S.\,G.  
        Automated Test Case Generation for Satellite FRD Using NLP and Large Language Model.  
        In \textit{2024 4th International Conference on Electrical, Computer, Communications and Mechatronics Engineering (ICECCME)}, 2024, pp.\,1–8.

  \item Sánchez Cuadrado, Jesús; Pérez-Soler, Sara; Guerra, Esther; De Lara, Juan.  
        Automating the Development of Task-oriented LLM-based Chatbots.  
        In \textit{Proceedings of the 6th ACM Conference on Conversational User Interfaces}, ACM, 2024.

  \item Arora, Chetan; Herda, Tomas; Homm, Verena.  
        Generating Test Scenarios from NL Requirements Using Retrieval-Augmented LLMs: An Industrial Study.  
        In \textit{2024 IEEE International Requirements Engineering Conference (RE)}, IEEE, 2024.

  \item Seifert, Daniel; Jöckel, Lisa; Trendowicz, Adam; Ciolkowski, Marcus; Honroth, Thorsten; Jedlitschka, Andreas.  
        Can Large Language Models (LLMs) Compete with Human Requirements Reviewers? – Replication of an Inspection Experiment on Requirements Documents.  
        \textit{Lecture Notes in Computer Science}, vol.\,15452 LNCS, Springer, 2025.

  \item Gohar, Usman; Hunter, Michael C.; Lutz, Robyn R.; Cohen, Myra B.  
        CoDefeater: Using LLMs to Find Defeaters in Assurance Cases.  
        In \textit{2024 39th IEEE/ACM International Conference on Automated Software Engineering (ASE)}, 2024, pp.\,2262–2267.

  \item Bai, Xingyuan; Huang, Shaobin; Wei, Chi; Wang, Rui.  
        Collaboration Between Intelligent Agents and Large Language Models: A Novel Approach for Enhancing Code Generation Capability.  
        \textit{Expert Systems with Applications}, vol.\,269, Art.~126357, 2025.

  \item Zhang, Li; Ouyang, Liubo; Xu, Zhuoqun.  
        Defect Correction Method for Software Requirements Text Using Large Language Models.  
        In \textit{2024 International Joint Conference on Neural Networks (IJCNN)}, 2024, pp.\,1–8.

  \item Bragilovski, Maxim; Van Can, Ashley T.; Dalpiaz, Fabiano; Sturm, Arnon.  
        Deriving Domain Models from User Stories: Human vs.\ Machines.  
        In \textit{2024 IEEE 32nd International Requirements Engineering Conference (RE)}, 2024, pp.\,31–42.

  \item De Vito, Gabriele; Palomba, Fabio; Gravino, Carmine; Di Martino, Sergio; Ferrucci, Filomena.  
        ECHO: An Approach to Enhance Use Case Quality Exploiting Large Language Models.  
        In \textit{2023 49th Euromicro Conference on Software Engineering and Advanced Applications (SEAA)}, 2023, pp.\,53–60.

  \item Ramasamy, Vijayalakshmi; Ramamoorthy, Suganya; Walia, Gursimran Singh; Kulpinski, Eli; Antreassian, Aaron.  
        Enhancing User Story Generation in Agile Software Development Through OpenAI and Prompt Engineering.  
        In \textit{2024 IEEE Frontiers in Education Conference (FIE)}, 2024, pp.\,1–8.

  \item Arulmohan, Sathurshan; Meurs, Marie-Jean; Mosser, Sébastien.  
        Extracting Domain Models from Textual Requirements in the Era of Large Language Models.  
        In \textit{2023 ACM/IEEE International Conference on Model Driven Engineering Languages and Systems Companion (MODELS-C)}, 2023, pp.\,580–587.

  \item Sami, Malik Abdul; Waseem, Muhammad; Zhang, Zheying; Rasheed, Zeeshan; Systä, Kari; Abrahamsson, Pekka.  
        Early Results of an AI Multi-Agent System for Requirements Elicitation and Analysis.  
        In \textit{Product-Focused Software Process Improvement}, Springer Nature Switzerland, 2025, pp.\,307–316.

  \item Zhao, Zelong; Zhang, Nan; Yu, Bin; Duan, Zhenhua.  
        Generating Java Code Pairing with ChatGPT.  
        \textit{Theoretical Computer Science}, vol.\,1021, Art.~114879, 2024.

  \item Görer, Binnur; Aydemir, Fatma Başak.  
        Generating Requirements Elicitation Interview Scripts with Large Language Models.  
        In \textit{Proceedings of the 31st IEEE International Requirements Engineering Conference Workshops (REW)}, IEEE, 2023, pp.\,44–51.

  \item Oswal, Jay U.; Kanakia, Harshil T.; Suktel, Devvrat.  
        Transforming Software Requirements into User Stories with GPT-3.5 – An AI-Powered Approach.  
        In \textit{2024 2nd International Conference on Intelligent Data Communication Technologies and Internet of Things (IDCIoT)}, IEEE, 2024, pp.\,913–920.

  \item Kotama, I. Nyoman Darma; Funabiki, Nobuo; Panduman, Yohanes Yohanie Fridelin; Brata, Komang Candra; Pradhana, Anak Agung Surya; Desnanjaya, I. Gusti Made Ngurah.  
        Implementation of Sensor Input Setup Assistance Service Using Generative AI for SEMAR IoT Application Server Platform.  
        \textit{INFORMATION}, vol.\,16, no.\,2, Feb.\ 2025.

  \item Wang, Yizhuo; Cui, Shiqi; Wan, Rongxin; Wang, Jingyi; Wang, Fanggang.  
        Large Language Models Based Communication Simulation Platform.  
        In \textit{2024 IEEE 24th International Conference on Communication Technology (ICCT)}, IEEE, 2024, pp.\,1891–1895.

  \item Blasek, Nico; Eichenmüller, Karl; Ernst, Bastian; Götz, Niklas; Nast, Benjamin; Sandkuhl, Kurt.  
        Large Language Models in Requirements Engineering for Digital Twins.  
        In \textit{CEUR Workshop Proceedings}, vol.\,3645, 2023.

  \item An, Zhiquan; Wan, Hongyan; Xiong, Teng; Wang, Bangchao.  
        LePB-SA4RE: A Lexicon-Enhanced and Prompt-Tuning BERT Model for Evolving Requirements Elicitation from App Reviews.  
        \textit{Applied Sciences – Basel}, vol.\,15, no.\,5, 2025.

  \item Fazelnia, Mohamad; Koscinski, Viktoria; Herzog, Spencer; Mirakhorli, Mehdi.  
        Lessons from the Use of Natural Language Inference (NLI) in Requirements Engineering Tasks.  
        In \textit{2024 IEEE 32nd International Requirements Engineering Conference (RE)}, IEEE, 2024, pp.\,103–115.

  \item Pereira, Juanan; López, Juan-Miguel; Garmendia, Xabier; Azanza, Maider.  
        Leveraging Open Source LLMs for Software Engineering Education and Training.  
        In \textit{2024 36th International Conference on Software Engineering Education and Training (CSEE\&T)}, IEEE, 2024, pp.\,1–10.

  \item Ferrari, Alessio; Abualhaija, Sallam; Arora, Chetan.  
        Model Generation with LLMs: From Requirements to UML Sequence Diagrams.  
        In \textit{2024 IEEE 32nd International Requirements Engineering Conference Workshops (REW)}, IEEE, 2024, pp.\,291–300.

  \item Zhao, Mengyan; Tao, Ran; Huang, Yanhong; Shi, Jianqi; Qin, Shengchao; Yang, Yang.  
        NL2CTL: Automatic Generation of Formal Requirements Specifications via Large Language Models.  
        In \textit{Lecture Notes in Computer Science (LNCS)}, vol.\,15394, Springer, 2024, pp.\,1–17.

  \item Chen, Boqi; Chen, Kua; Hassani, Shabnam; Yang, Yujing; Amyot, Daniel; Lessard, Lysanne; Mussbacher, Gunter; Sabetzadeh, Mehrdad; Varró, Dániel.  
        On the Use of GPT-4 for Creating Goal Models: An Exploratory Study.  
        In \textit{Proceedings of the IEEE 31st International Requirements Engineering Conference Workshops (REW)}, IEEE, 2023, pp.\,262–271.

  \item Sivakumar, Mithila; Belle, Alvine B.; Shan, Jinjun; Shahandashti, Kimya Khakzad.  
        Prompting GPT-4 to Support Automatic Safety Case Generation.  
        In \textit{Expert Systems with Applications}, vol.\,255, Elsevier, 2024, p.\,124653.

  \item Ruan, Kun; Chen, Xiaohong; Jin, Zhi.  
        Requirements Modeling Aided by ChatGPT: An Experience in Embedded Systems.  
        In \textit{Proceedings of the IEEE 31st International Requirements Engineering Conference Workshops (REW)}, IEEE, 2023.

  \item Santos, Sarah; Breaux, Travis; Norton, Thomas; Haghighi, Sara; Ghanavati, Sepideh.  
        Requirements Satisfiability with In-Context Learning.  
        In \textit{2024 IEEE International Requirements Engineering Conference (RE)}, IEEE, 2024.

  \item Luitel, Dipeeka; Nejati, Shiva; Sabetzadeh, Mehrdad.  
        Requirements-Driven Slicing of Simulink Models Using LLMs.  
        In \textit{Proceedings of the IEEE 32nd International Requirements Engineering Conference Workshops (REW)}, IEEE, 2024, pp.\,72–82.

  \item Li, Yishu; Keung, Jacky; Yang, Zhen; Ma, Xiaoxue; Zhang, Jingyu; Liu, Shuo.  
        SimAC: Simulating Agile Collaboration to Generate Acceptance Criteria in User Story Elaboration.  
        \textit{Automated Software Engineering}, vol.\,31, no.\,2, Nov.\ 2024.

  \item Preda, Anamaria-Roberta; Mayr-Dorn, Christoph; Mashkoor, Atif; Egyed, Alexander.  
        Supporting High-Level to Low-Level Requirements Coverage Reviewing with Large Language Models.  
        In \textit{Proceedings of the IEEE/ACM 21st International Conference on Mining Software Repositories (MSR)}, IEEE/ACM, 2024, pp.\,242–253.

  \item Rahman, Tajmilur; Zhu, Yuecai; Maha, Lamyea; Roy, Chanchal; Roy, Banani; Schneider, Kevin.  
        Take Loads Off Your Developers: Automated User Story Generation Using Large Language Model.  
        In \textit{Proceedings of the IEEE International Conference on Software Maintenance and Evolution (ICSME)}, IEEE, 2024.
\end{prlist}

\end{document}